# EVIDENCE FOR HARD CHIRAL LOGARITHMS IN QUENCHED LATTICE QCD [†]

S. Kim and D. K. Sinclair

*HEP Division,*
*Argonne National Laboratory, 9700 South Cass Avenue, Argonne, IL 60439*

## ABSTRACT

We present the first direct evidence that quenched QCD differs from full QCD in the chiral ($m_q \to 0$) limit, as predicted by chiral perturbation theory, from our quenched lattice QCD simulations at $\beta = 6/g^2 = 6.0$. We measured the spectrum of light hadrons on $16^3 \times 64$, $24^3 \times 64$ and $32^3 \times 64$, using staggered quarks of masses $m_q = 0.01$, $m_q = 0.005$ and $m_q = 0.0025$. The pion masses showed clear evidence for logarithmic violations of the PCAC relation $m_\pi^2 \propto m_q$, as predicted by quenched chiral perturbation theory. The dependence on spatial lattice volume precludes this being a finite size effect. No evidence was seen for such chiral logarithms in the behaviour of the chiral condensate $\langle \bar\psi\psi \rangle$.

---

[†]Work supported by the U.S. Department of Energy, Division of High Energy Physics, Contract W-31-109-ENG-38

# I. INTRODUCTION

Simulation of full lattice QCD with light quarks, turns out to be very time consuming. Most of the computer time involved in such simulations is spent in calculating (changes in) the determinant of the Dirac operator. For this reason, calculations on the largest lattices are usually performed using the quenched (valence-quark) approximation where this determinant is replaced by unity in the generation of gauge field configurations. The questions, how reliable is this approximation and where does it break down, naturally arise. Clearly the quenched approximation does not give the correct running of the coupling constant. On the other hand, hadron spectra and related quantities, calculated with u and d quark masses which are unphysically large, show good agreement between quenched and full QCD.

However, quenched QCD clearly fails in its treatment of the flavour singlet pseudoscalar meson – the $\eta'$. In full QCD, the $\eta'$ propagator can be represented by the sum

$$\text{———} \ + \ \text{⊃ ⊂} \ + \ \text{⊃ ◯ ⊂} \ + \cdots \tag{1}$$

The first term is just the pion propagator and has a simple pole at $m_\pi^2$. Summing this geometric series shifts this pole to $m_{\eta'}^2$. In the quenched case, only the first two terms survive. This is due to the absence of closed fermion loops in the quenched approximation. Thus the propagator has a simple pole from the first term and a double pole from the second, both at $m_\pi^2$, an unphysical behaviour. In fact, since the second term includes a factor of $N_f$ (the number of flavours), taking the limit $N_f \to 0$ as is appropriate for the quenched approximation removes this second term in equation 1. Unfortunately, this contribution, with its double pole, reappears in $\eta'$ loop contributions. To study how this feeds back into the other propagators and observables of the quenched theory, Sharpe [1,2], and Bernard and Golterman [3,4] developed quenched chiral perturbation theory. The presence of the double pole in the $\eta'$ propagator, gives a logarithmic infrared divergence to loops involving the $\eta'$ propagator in the chiral limit of this effective field theory. This means that amplitudes will develop spurious powers of $\ln(m_\pi^2/\Lambda^2)$. These chiral logarithms are "hard" in that they control the chiral limit of the theory. This distinguishes them from the chiral logarithms of full QCD, which are "soft", that is they are multiplied by (powers of) $m_\pi^2$, and so do not contribute in the chiral limit. Thus quenched QCD should produce results which differ from full QCD, at least in the limit $m_\pi \to 0$. It is important to know how large these effects are at quark masses small enough to give the correct pion mass, or at least small enough to be able to extrapolate to the correct pion mass, since these effects represent systematic errors introduced by the quenched approximation.

Prior to our work, no evidence for such chiral logarithms had been reported from simulations of quenched QCD. We report on our measurement of pion masses and $\langle\bar{\psi}\psi\rangle$ on $16^3 \times 64$, $24^3 \times 64$ and $32^3 \times 64$ lattices at $\beta = 6.0$ at staggered quark masses $m_q = 0.01$, $m_q = 0.005$ and $m_q = 0.0025$ (in lattice units). A complete report of hadron masses, and finite size effects from these simulations will be given elsewhere [5]. At these low quark masses chiral logarithms are evident in the pion mass data, but not directly in the $\langle\bar{\psi}\psi\rangle$ data. Kuramashi, et al. [6] had noticed earlier that our preliminary data for the pion mass from a subset of our $32^3 \times 64$ lattice configurations [7], appeared to show evidence for chiral logarithms. We have now completed the finite size analysis needed to rule out this being a finite size effect.

In section 2 we describe our simulations and measurements, and present our results. These results are discussed in section 3.

## II. SIMULATIONS AND MEASUREMENTS

Quenched gauge field configurations with $\beta = 6.0$ were generated alternating one 10 hit metropolis sweep with one overrelaxation sweep. On each lattice, configurations were separated by 1000 sweeps after discarding the first 20,000 sweeps for equilibration. We have generated and analysed 200 configurations on a $32^3 \times 64$ lattice, 339 configurations on a $24^3 \times 64$ lattice, and 410 configurations on a $16^3 \times 64$ lattice, allowing us to estimate finite size effects. For meson spectroscopy we used a "corner" wall source on time-slice 1 to generate our quark propagators. On the $16^3 \times 64$ lattice we also used a second "corner" wall source on time-slice 33. A stochastic estimator for $\langle\bar{\psi}\psi\rangle$ was calculated for each configuration using a single gaussian-noise source. We refer the reader to our earlier work at $\beta = 6.5$ for details of our methods [8].

The goldstone pion propagators are fit to the form

$$C_\pi(T) = A\{exp[-m_\pi T] + exp[-m_\pi(N_t - T)]\} \tag{2}$$

while the local $\pi_2$-$f_0$ propagators are fit to the form

$$C_{\pi_2,f_0}(T) = B\{exp[-m_{\pi_2}T] + exp[-m_{\pi_2}(N_t - T)]\} + (-1)^T\{exp[-m_{f_0}T] + exp[-m_{f_0}(N_t - T)]\} \tag{3}$$

where $T$ is the difference in times between the timeslice containing the source and that of the sink, $N_t$ is the temporal extent of the lattice and $\pi_2$ is the local non-goldstone pion. Correlated fits are used in each case. The quantity

$$quality \equiv \frac{confidence\ level \times degrees\ of\ freedom}{error}, \quad (4)$$

introduced by the HEMCGC collaboration [9], is used as a guide when choosing among fits of acceptable $\chi^2$/confidence level. This is merely an attempt to quantify our prejudice that good fits should not only have a high confidence level, but should also cover a large range of T values, and have small errors. Table I gives values of $m_\pi$, and table II gives values of $m_{\pi_2}$ from these simulations, as a function of $m_q$ and lattice size.

For full QCD $m_\pi^2 = 2\mu m_q$, as $m_q \to 0$ where $\mu$ is a constant with the dimensions of a mass. In Fig. 1 we have plotted $m_\pi^2/m_q$ against $m_q$ (both in lattice units) from our simulations. We notice that there are departures from the constant value predicted for full QCD. Moreover, while the values for the $16^3 \times 64$ lattice lie considerably above the corresponding numbers for the larger lattices, the $24^3 \times 64$ and $32^3 \times 64$ are consistent with one another. Hence this departure from current algebra does not appear to be a finite size effect for the two larger lattices. To check that this departure is statistically significant, we have attempted a correlated fit of the form $m_\pi^2 = 2\mu m_q$ to our 3 data points for the $32^3 \times 64$ lattice and find a $\chi^2$ per degree of freedom of $> 80$ which rules out this simple relation.

For quenched chiral perturbation theory, Sharpe [2,10] has summed the leading chiral logarithms to predict a form

$$m_\pi^2 = 2\mu m_q (m_{\eta'}^2/\Lambda^2)^{-\delta} \quad (5)$$

where $m_{\eta'} = m_\pi$. We have therefore attempted to fit our lattice results to the form

$$\ln m_\pi^2 = c + \ln m_q - \delta \ln m_{\eta'}^2 \quad (6)$$

For staggered quarks at $\beta = 6.0$, the goldstone and non-goldstone pions are not degenerate. The non-goldstone pions, including the $\eta'$, are, however, much closer to being degenerate [11,12]. We therefore take $m_{\eta'} = m_{\pi_2}$ in our fits to equation 6. The fit to the $32^3 \times 64$ data yields $c = 1.62(1)$ and $\delta = 0.056(5)$, with a confidence level of 38%. (For comparison, if we take $m_{\eta'} = m_\pi$ we get $c = 1.645(8)$ and $\delta = 0.039(3)$ with a confidence level of 48%.) We have also fit our data to the form

$$\ln m_\pi^2 = c + \ln m_q - \delta \ln m_{\eta'}^2 + dm_\pi^2 \quad (7)$$

to parameterize departures from the chiral limit [13]. In the case $m_{\eta'} = m_{\pi_2}$ this yields $c = 1.5(1)$, $\delta = 0.09(4)$ and $d = 0.6(7)$ which, as we could have guessed from the good fits obtained to the form in equation 6, indicates that the introduction of the extra parameter $d$ is not justified. Finally, we have attempted to fit to the form

$$m_\pi^2/m_q = C + am_\pi^2 \tag{8}$$

to see if the departure from standard PCAC can be described as an $\mathcal{O}(m_\pi^2)$ correction. Here, the $\chi^2$ per degree of freedom of the fit $\approx 6$ (confidence level $< 2\%$). Thus the chiral logarithm interpretation is favoured.

For completeness, in Fig. 3 we give an effective mass plot for $m_\pi^2/m_q$ for our $32^3 \times 64$ and $24^3 \times 64$ lattices. Here we wish to indicate that the effect we are seeing is not an artifice of the systematic errors introduced by our particular choice of fits to the goldstone pion propagator. While using such an expanded scale and squaring the pion mass exaggerates the fluctuations and tends to obscure signs of a plateau, one property is clear; there is a statistically significant ordering of this quantity, with the $m_q = 0.01$ values lying lowest, and the $m_q = 0.0025$ lying highest. This supports the behaviour observed in Fig. 1, which forms the basis of our evidence for quenched chiral logarithms.

Let us now turn to the consideration of the chiral condensate, $\langle \bar{\psi}\psi \rangle$. In full QCD $\langle \bar{\psi}\psi \rangle$ approaches a constant as $m_q \to 0$. In fact $\langle \bar{\psi}\psi \rangle = \mu f_\pi^2$ (where the continuum $\langle \bar{\psi}\psi \rangle$ is for a single flavour). This leads to the Gell-Mann–Oakes–Renner relation $f_\pi^2 m_\pi^2 = 2m_q \langle \bar{\psi}\psi \rangle$ which remains valid in the quenched case. For quenched QCD, $f_\pi$ remains free of chiral logarithms, at least in the leading-logarithmic approximation [1–4,10]. Hence $\langle \bar{\psi}\psi \rangle$ behaves as

$$\langle \bar{\psi}\psi \rangle = \mu f_\pi^2 (m_{\eta'}^2/\Lambda^2)^{-\delta} \tag{9}$$

with $f_\pi \to$ constant, as $m_q \to 0$. There exists, however, an additional complication. $\langle \bar{\psi}\psi \rangle$ as traditionally defined (at least on the lattice) has a quadratic ultraviolet divergence. The above formulae are only true only for an appropriately subtracted $\langle \bar{\psi}\psi \rangle$. For full QCD this means our lattice $\langle \bar{\psi}\psi \rangle$ should be fit to the form

$$\langle \bar{\psi}\psi \rangle = c(1 + am_\pi^2), \tag{10}$$

while the quenched lattice value should be fit to

$$\langle \bar{\psi}\psi \rangle = c(1 + am_\pi^2)(m_{\eta'}^2)^{-\delta} \tag{11}$$

where in both cases $a$ has a piece proportional to the square of the inverse lattice spacing, which is the lattice regulated ultraviolet divergence. (That the quadratic divergence appears in a simple multiplicative factor in equation 11 was shown by Kilcup and Sharpe [14].) Our measured $\langle \bar{\psi}\psi \rangle$ values are given in table III and plotted in Fig. 2. (Note that, our lattice $\langle\bar{\psi}\psi\rangle$ is normalized to 4 quark flavours, the natural normalization with staggered fermions.) We have fitted $\langle\bar{\psi}\psi\rangle$ to the form in equation 10 and found $c = 0.0296(2)$ and $a = 32.7(4)$, with a confidence level of 74%. This is such a good fit that we conclude that there is no evidence for chiral logarithms in $\langle\bar{\psi}\psi\rangle$, other than those in $m_\pi$. (Because of this we should not expect a good fit if we replace $m_\pi^2$ with $m_q$. Indeed, such a fit has a confidence level of only 5%.)

Finally we have reanalysed the data from our $\beta = 6.5$ simulations [8]. Here, although $m_\pi^2/m_q$, does show some dependence on $m_q$, it does not appear to be statistically significant. We were able to fit this ratio to a constant $c = 2.55(4)$, with a confidence level of 23%. Attempts at fits with chiral logarithms or a term linear in $m_\pi^2$ yielded even worse fits.

### III. DISCUSSIONS AND CONCLUSIONS

We have observed clear evidence for quenched chiral logarithms in the pion spectrum of quenched lattice QCD at $\beta = 6.0$. The reason we were able to observe this effect appears to be that, by using a large ($32^3 \times 64$) lattice, we were able to accommodate smaller quark masses (as low as $m_q = 0.0025$) than had been considered previously. By measuring these masses on $16^3 \times 64$ and $24^3 \times 64$ lattices in addition to the $32^3 \times 64$ lattice, we have demonstrated that the pion mass has negligible finite size effects for spatial boxes of length $\geq 24$, at these quark masses. Our fits were to the leading-logarithmic expressions of Sharpe [2,10], rather than the 1-loop calculations [1,3,4]. The reason for this choice is simple: because the leading-logarithmic sum comes from a renormalization-group argument, the arbitrariness of the scale $\Lambda$ is obvious, and the value of the exponent $\delta$ is independent of $\Lambda$. This is not true of the 1-loop result.

Our calculated value of $\delta$ ($\delta = 0.056(5)$) is considerably below the estimated value ($\delta \approx 0.2$). Gupta [13], who has combined our results with those of his group [15] (which show no evidence for quenched chiral logarithms), finds $\delta \approx 0.13$. The difference is the inclusion of corrections $\mathcal{O}(m_\pi^2)$ and $\mathcal{O}(m_\pi^4)$, which tend to cancel the contribution of the logarithm, allowing a smooth matching with the higher mass results. We also saw some indication of this effect when we included an $\mathcal{O}(m_\pi^2)$ contribution, leading to $\delta \approx 0.09$. However, in our case this difference was not significant. If $\delta \approx 0.2$, then it would appear that the corrections which are a power series in $m_\pi^2$ are "conspiring" to mask the chiral logarithms down to rather small quark masses. Let us try to estimate the size of the errors in the pion mass incurred by using the quenched approximation, and coming from quenched chiral logarithms. In the spirit of Gupta, let us assume that at $m_q = 0.01$, this error is negligible. From table 2 we see that the error in the pion mass at $m_q = 0.0025$ is $\approx 2.5\%$. Taking the inverse lattice spacing to be $\approx 2 GeV$, assuming an effective $\delta$ of 0.1 from $m_q = 0.0025$ down to the physical pion mass, and using $m_{\eta'} = m_\pi$, we find a further error of $\approx 6\%$ in going to the physical pion mass. If we had used $m_{\eta'} = m_{\pi_2}$ this error would have been even less. Thus we would expect that the quenched chiral logarithms will introduce an error of 5 – 10% in the pion mass. Of course, at $\beta = 6.0$, this is a considerably smaller source of systematic errors in the pion mass than the flavour symmetry breaking between the goldstone and non-goldstone pions.

We have observed no evidence for quenched chiral logarithms in the chiral condensate $\langle \bar\psi\psi \rangle$. At least part of the problem here could be the presence of a large $\mathcal{O}(m_\pi^2)$ term, due to the fact that our definition of $\langle \bar\psi\psi \rangle$ is ultraviolet divergent for finite $m_q$. This term tends to mask any other dependence on $m_q$. In addition, as we have seen in the case of the pion mass, power series in $m_\pi^2$ can mask chiral logarithms down to quite low quark masses. There is no reason, a priori, why the chiral logarithms in $\langle \bar\psi\psi \rangle$ should not be suppressed to even lower quark masses than is the case for $m_\pi$, by the power series behaviour of $f_\pi$.

Our measurements at $\beta = 6.5$, which do not probe as far towards the chiral limit as those at $\beta = 6.0$, do not show signs of quenched chiral logarithms. This is unfortunate, since at $\beta = 6.5$, the pion masses show little sign of flavour symmetry breaking, so the systematic errors due to lattice breaking of continuum symmetries are small. Thus had we observed any anomalous behaviour, we could have been confident that it was a property of the continuum theory.

We conclude, therefore, that anomalous chiral logarithms do appear in at least some physical quantities in quenched QCD. Clearly more work is needed. The effects we have seen are small, and it is hoped that some scheme similar to the one used by Kuramashi et al. [6] for the $\eta'$ mass, in which the effects of the missing loops are estimated, could be applied to more general amplitudes in quenched QCD.

**Acknowledgements**

One of us, S.K., would like to thank A. Ukawa for suggesting to him that our preliminary results might display chiral logarithms. We also wish to thank our colleagues G. T. Bodwin and A. W. White for helpful discussions about the ultra-violet divergences of $\langle \bar\psi\psi \rangle$. This research was conducted using the Intel Touchstone Delta and the two Intel Paragon computers operated by Caltech on behalf of the Concurrent Supercomputer Consortium. We thank R. L. Stevens and W. D. Gropp of the MCS division and F. Y. Fradin, the associate Laboratory Director for Physical Research at Argonne National Laboratory for allowing us access to these CCSF computers. We also wish to thank the CCSF staff at Caltech for their support.

|  | $16^3 \times 64$ | | $24^3 \times 64$ | | $32^3 \times 64$ | |
|---|---|---|---|---|---|---|
| $m_q$ | range | $m_\pi$ | range | $m_\pi$ | range | $m_\pi$ |
| 0.0100 | 9–30 | 0.2414(4) | 13–30 | 0.2404(3) | 16–30 | 0.2406(3) |
| 0.0050 | 5–30 | 0.1743(4) | 11–30 | 0.1720(3) | 16–31 | 0.1723(3) |
| 0.0025 | 4–30 | 0.1265(4) | 11–30 | 0.1232(3) | 18–31 | 0.1233(4) |

TABLE I. $m_\pi$, at $\beta = 6.0$

|  | $16^3 \times 64$ | | $24^3 \times 64$ | | $32^3 \times 64$ | |
|---|---|---|---|---|---|---|
| $m_q$ | range | $m_{\pi_2}$ | range | $m_{\pi_2}$ | range | $m_{\pi_2}$ |
| 0.0100 | 2–32 | 0.291(2) | 10–32 | 0.296(2) | 6–25 | 0.291(1) |
| 0.0050 | 2–32 | 0.236(3) | 5–32 | 0.233(2) | 4–19 | 0.232(2) |
| 0.0025 | 0–23 | 0.203(5) | 2–29 | 0.195(3) | 4–23 | 0.195(3) |

TABLE II. $m_{\pi_2}$, at $\beta = 6.0$

|  | $16^3 \times 64$ | $24^3 \times 64$ | $32^3 \times 64$ |
|---|---|---|---|
| $m_q$ | $\langle\bar\psi\psi\rangle$ | $\langle\bar\psi\psi\rangle$ | $\langle\bar\psi\psi\rangle$ |
| 0.0100 | 0.08541(19) | 0.08566(11) | 0.08552(11) |
| 0.0050 | 0.05856(27) | 0.05818(15) | 0.05832(12) |
| 0.0025 | 0.04426(34) | 0.04411(21) | 0.04424(18) |

TABLE III. $\langle\bar\psi\psi\rangle$, at $\beta = 6.0$

**Figure captions**

1. $m_\pi^2/m_q$ as a function of $m_{\pi_2}$. This is a log-log plot, so that the predicted power law behaviour is a straight line. The solid line is the fit to equation 6.

2. $\langle\bar\psi\psi\rangle$ as a function of $m_{\pi_2}^2$. The solid line is the fit to equation 10.

3. $m_\pi^2(effective)/m_q$ as a function of $T_{min}$: a) On a $32^3 \times 64$ lattice. b) On a $24^3 \times 64$ lattice. The solid lines show the fits of table I; the dashed lines give their errors.

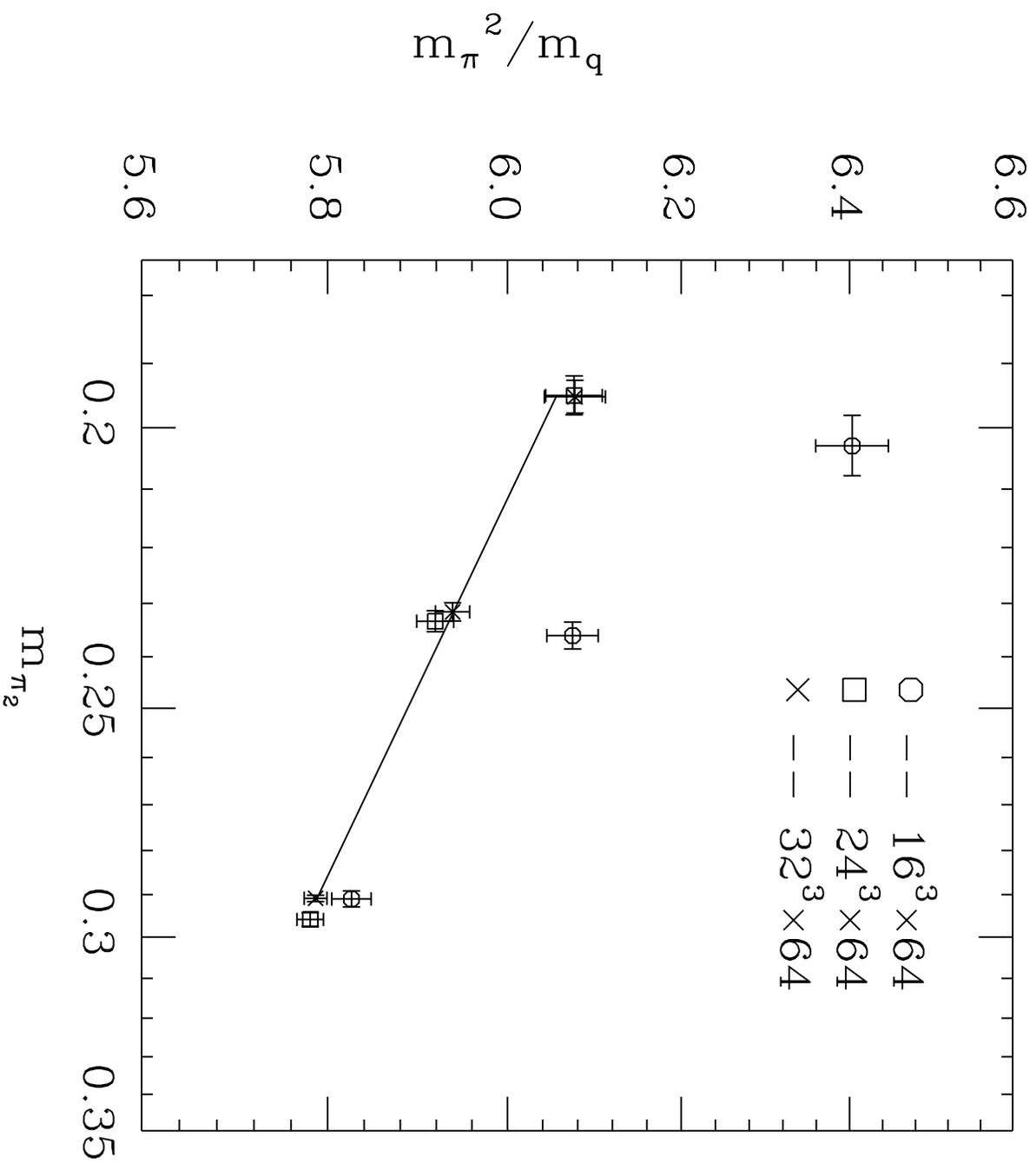

Fig. 1

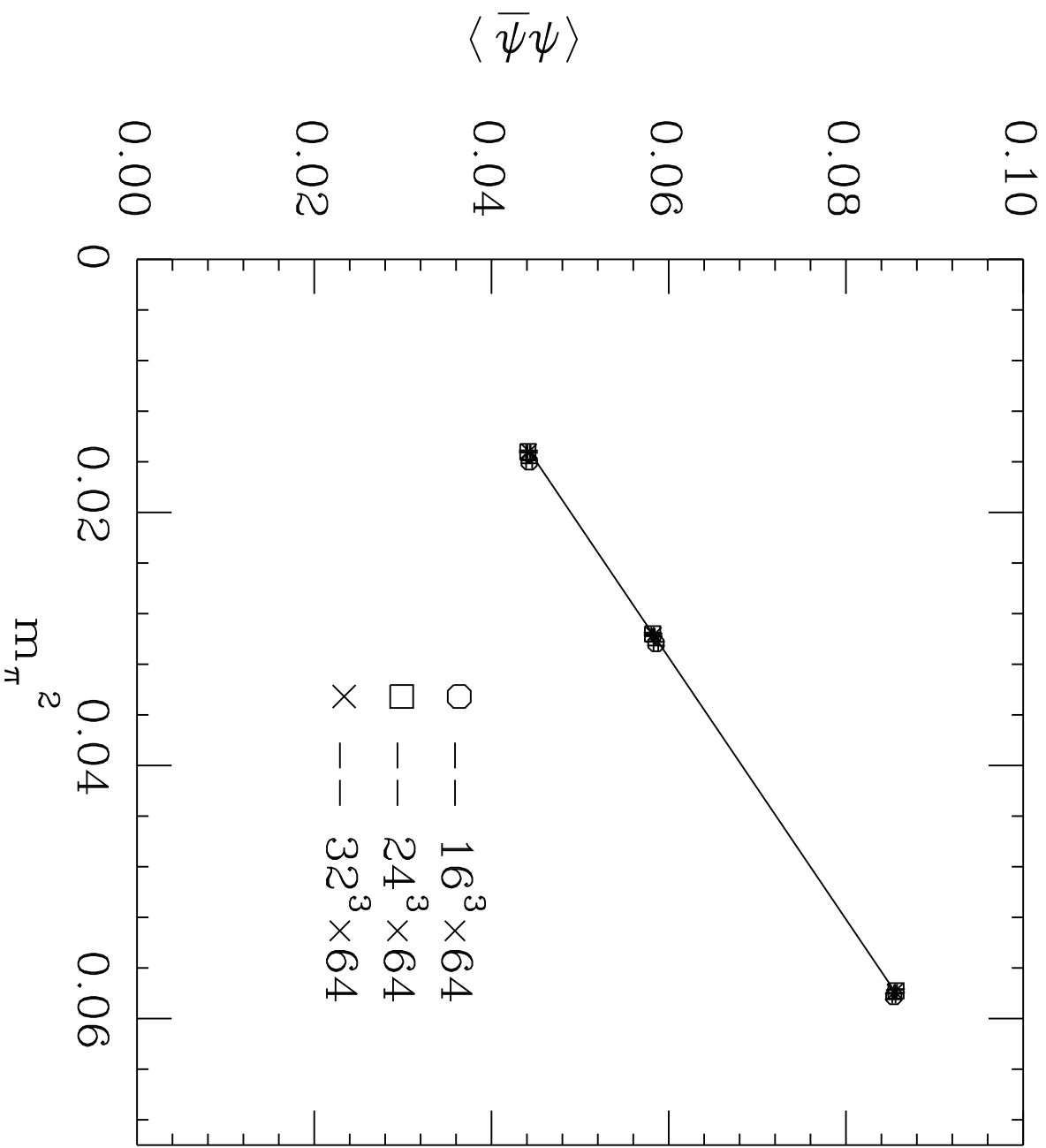

Fig. 2

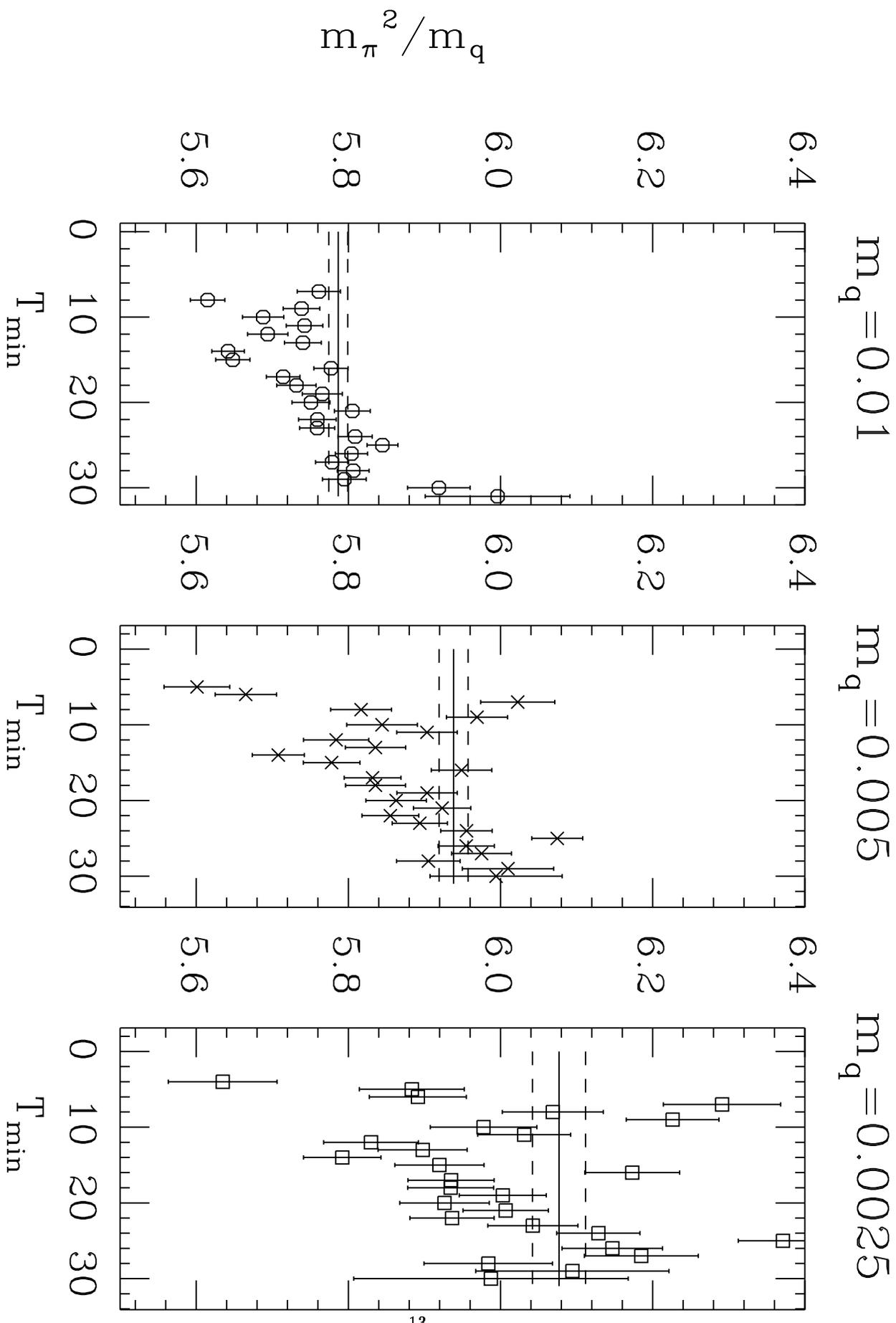

Fig. 3a

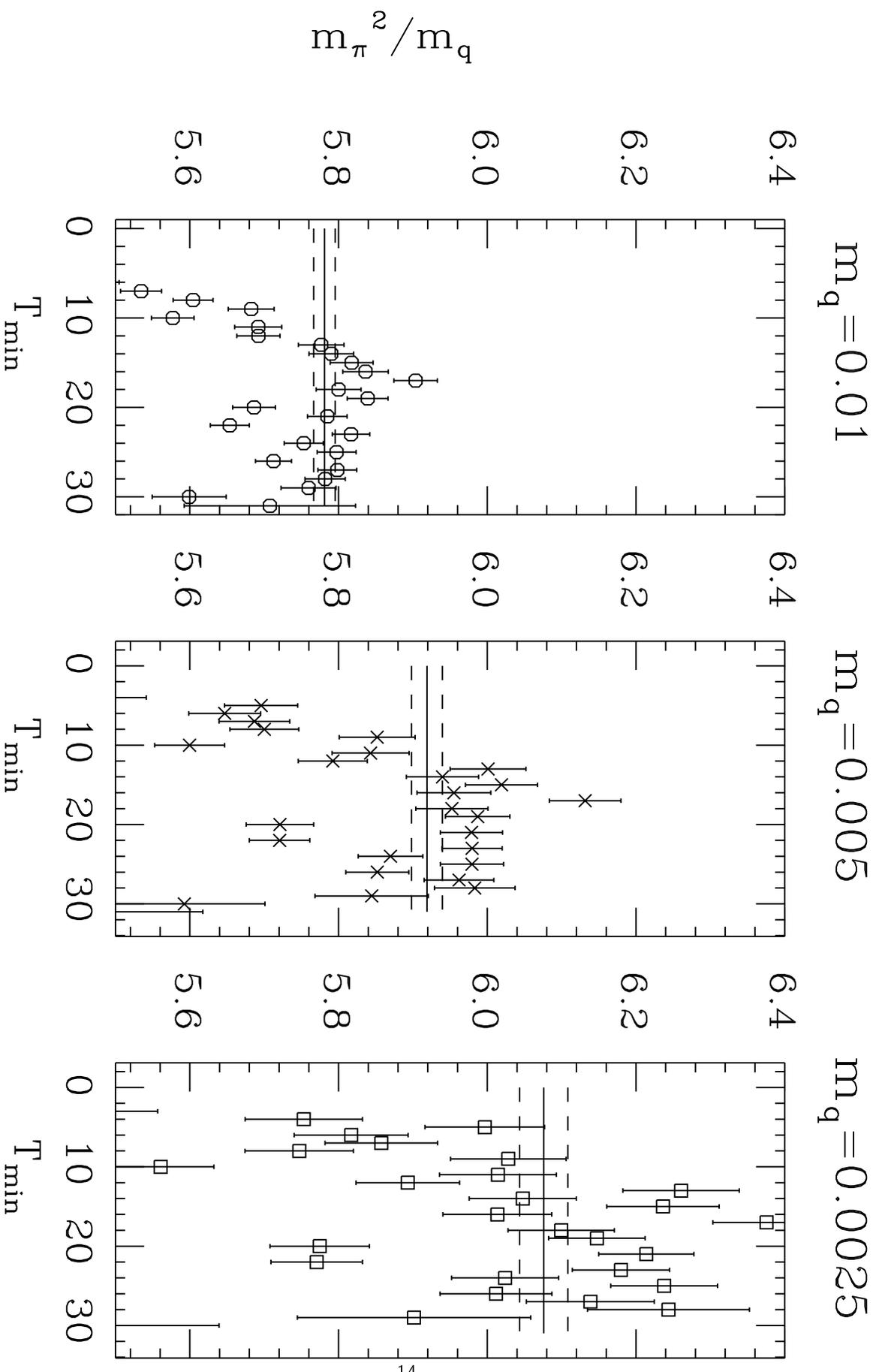

Fig. 3b

14